# Some Novel Results From Analysis of Move To Front (MTF) List Accessing Algorithm

[1]Rakesh Mohanty, [2]Shiba Prasad Dash, [3]Burle Sharma, [4]Sangita Patel

[1]Department of Computer Science and Engineering, Indian Institute of Technology, Madras, Chennai, India-600036

[2,3]Department of Computer Science and Engineering, Veer Surendra Sai University of Technology, Burla, Sambalpur, Odisha, India

[4]Department of Computer Science Sambalpur University Institute of Information Technology Burla, Sambalpur, Odisha, India

*Abstract* - List accessing problem has been studied as a problem of significant theoretical and practical interest in the context of linear search. Various list accessing algorithms have been proposed in the literature and their performances have been analyzed theoretically and experimentally. Move-To-Front (MTF), Transpose (TRANS) and Frequency Count (FC) are the three primitive and widely used list accessing algorithms. Most of the other list accessing algorithms are the variants of these three algorithms. As mentioned in the literature as an open problem, direct bounds on the behavior and performance of these list accessing algorithms are needed to allow realistic comparisons. MTF has been proved to be the best performing online algorithm till date in the literature for real life inputs with locality of reference. Motivated by the above challenging research issue, in this paper we have generated four types of input request sequences corresponding to real life inputs without locality of reference. Using these types of request sequences, we have made an analytical study for evaluating the performance of MTF list accessing algorithm to obtain some novel and interesting theoretical results.

*Keywords:* Algorithm; Data Structure; Linked List; Linear Search; List Accessing; Move-To-Front.

## I. INTRODUCTION

*Linear Search* is a basic and simple search method for unsorted linear list. The *List Update Problem (LUP)* is one of the popular problems which have been studied extensively for the last few decades in the context of linear search. The input to the LUP is a list of distinct items and a sequence of requests. Each *request* corresponds to an operation on an item of the list. The request may be either an access or an insert or a delete operation. Since insert and delete operations are special case of access operation, we consider only access operation for simplicity and hence the problem is also known as *List Accessing Problem (LAP)*. When a request from a request sequence is served on the list, the requested item is accessed in the list by incurring some access cost using a *cost model*. After accessing the requested item, the list is reorganized so that the frequently accessed items are moved towards the front of the list to reduce the future access cost. In the LAP, our goal is to obtain the optimal access cost by efficiently reorganizing the list while serving a request sequence on the list.

*a) List Accessing Cost Models*
Most widely used cost models for LAP are *full cost model* and *partial cost model*. The full cost model is popularly known standard cost model defined by Sleator and Tarjan [5]. In this model the access cost for a requested item is the position of the item from the front of the list. So the cost of accessing the *ith* item in the list is *i*. Immediately after an access, the accessed item can be moved anywhere towards the front of the list without paying any cost. This type of exchange is called a *free exchange*. Any other exchange of two adjacent items in the list costs 1. This type of exchange is known as *paid exchange*. In partial cost model the cost of accessing an item is the number of comparisons made with the preceding item in the list before accessing the item. The access cost of *ith* item in the list is *i-1*, since it requires *i-1* comparisons before accessing the item *i*. The reorganization cost is the minimum number of paid exchanges. So the total cost is the sum of the access cost and the reorganization cost.

*b) Primitive List Accessing Algorithms*
An algorithm which reorganizes the list and minimizes the reorganization and access cost while serving a request sequence is called a *list accessing algorithm*. Various list accessing algorithms have been proposed in the literature. All the list accessing algorithms developed are the variants of the following three primitive algorithms.

*Move-To-Front (MTF):* After accessing an item *x* in the list, *x* is immediately moved to the front of the list.

*Transpose (TRANS):* After accessing an item *x* in the list, *x* is immediately moved forward one position in the list by exchanging it with the immediately preceding item.

*Frequency Count (FC):* A frequency counter is maintained for each of the items of the list as per the number of occurrences of each item in the request sequence. When an item is accessed in the list from a request sequence, the corresponding frequency counter is increased by one. The list is reorganized and maintained in non increasing order of the access frequencies at any instant of time.

*c) Applications and Motivation*
List accessing algorithms are extensively used for data compression [7]. Some other popular applications of list accessing algorithms are maintaining small dictionaries, organizing the list of identifiers maintained by compilers and interpreters, resolving collisions in a hash table, computing point maxima and convex hulls in computational geometry.

The majority of the literature deals with analysis of various list accessing algorithms without consideration of any specific types of request sequences corresponding to real life applications. MTF has been proved to be the best performing online algorithm till date in the literature for real life inputs with locality of reference [12]. One interesting aspect of research is to generate some new special types of request sequences without locality of reference which correspond to some real life inputs. Analyzing the performance of MTF algorithm using these request sequences has a practical significance. In this work, we have considered four special types of request sequence without locality of reference. Our objective is to evaluate the performance of MTF algorithm based on access cost for these four types of request sequences.

*d) Literature Review*

Study of list accessing problem was initiated by the pioneering work of McCabe [1] in 1965 with introduction of two algorithms Move-To-Front (MTF) and Transpose (TRANS). From 1965 to 1985, the list accessing problem was studied by many researchers under the assumption that an input request sequence is generated by a probability distribution as mentioned in [2] and [3]. Hester and Hirschberg [4] have provided an extensive survey on average case analysis of list accessing algorithms. Sleator and Tarjan [5] in 1985 in their seminal paper have made a competitive analysis of MTF list accessing algorithm. The first use of randomization and the demonstration of its advantage in the competitive analysis were done by Borodin, Linial and Saks [6] with respect to material task system in 1985. Albers [10] in 1994 introduced the concept of look ahead in the list accessing problem. Bachrach and et.al. [8] have provided an extensive theoretical and experimental study of online list accessing algorithm in 2002. The study of locality of reference in list accessing problem was initiated by Angelopoulos [12] in 2006, where he proved MTF is superior to all algorithms. Relatively less work has been done on the offline algorithms for the list accessing problem. Reingold and Westbrook [9] in 1996 have developed an optimum offline list accessing algorithm by using subset transfer theorem that takes $\theta(2^n mn!)$ time. Ambuhl [11] in 2000 has proved that offline list update is NP-hard. A survey of important theoretical and experimental results related to on-line algorithms for list accessing problem is done in [13]. A classification of request sequences and few analytical results for MTF algorithm have been mentioned in [14].

*e) Our Contribution*

In our work we have generated some special patterns of request sequences corresponding to some real life inputs without locality of reference. Using these specific types of request sequences we have performed an analytical study of MTF list accessing algorithm. We have obtained some novel interesting theoretical results for evaluation of performance of MTF algorithm. We have represented the access cost of MTF algorithm for two specific types of request sequences in a graph using our analytical results.

*f) Organization of the paper*

The paper is organized as follows. Introduction and literature review is presented in section I. Section II contains some novel analytical results for MTF algorithm. Section III provides the concluding remarks and scope for future research work.

## II. NOVEL ANALYTICAL RESULTS FOR MTF WITH SPECIAL TYPES OF REQUEST SEQUENCES

A request sequence has *locality of reference*, if a single item or a small group of adjacent items from the list are frequently repeated in the request sequence. In many real life applications the request sequence consists of one or more repetitions of different configurations of the whole list. These request sequences do not have locality of reference property. In our work, we have considered some special types of request sequences that are repetitions of the same permutation of the list. Let $\ell = <\ell_1, \ell_2, \ell_3 \ldots \ell_n>$ be an unsorted list of items. Let $\sigma = <\sigma_1, \sigma_2, \sigma_3, \ldots, \sigma_m>$ be a request sequence of size $m$ such that $\sigma_i \epsilon$ for $i = 1, 2, 3, m$. For each item in the list, the list accessing algorithm must serve the request $\sigma_i$ in the order of its arrival. Let $k \geq 1$ be a positive integer that specifies the number of times a particular permutation of the list $\ell$ is repeated in the request sequence $\sigma$.

*a) Special Types of Request Sequences*

We have considered the following four types of request sequences for our analysis.

*Type1 ($T_1$)*- Let $\Pi_\ell = <\ell_1, \ell_2, \ell_3, \ldots, \ell_n>$ be a permutation of the list $\ell$ that consists of all the items of the list in the same order as in the list. $T_1 = \sigma = \Pi_\ell, \Pi_\ell, \ldots$(k times) $= (\Pi_\ell)$.

*Type2 ($T_2$)*- Let $\Pi_r = <\ell_n, \ell_{n-1}, \ldots, \ell_2, \ell_1>$ be a permutation of the list $\ell$ that consists of all the items of the list in the reverse order of the list. $T_2 = \sigma = \Pi_r, \Pi_r, \ldots$(k times) $=(\Pi_r)$

*Type3 ($T_3$)*- Let $\Pi_x$ be any permutation of the list $\ell$ except $\Pi_\ell$ and $\Pi_r$. $T_3 = \sigma = \Pi_x, \Pi_x, \ldots$(k times) $= (\Pi_x)\backslash$

*Type4 ($T_4$)*- Let $\delta_x$ be any subsequence of the list $\ell$ of size $q$ such that $q < n$. $T_4 = \sigma = \delta_x, \delta_x, \ldots, \delta_x$ (k times) $= (\delta_x)^k$.

*b) Novel Analytical Results*

We have obtained some interesting theoretical results for computing the access cost of MTF algorithm for specific types of request sequences like $T_1, T_2, T_3$ and $T_4$. Here we consider the Full Cost Model and Singly Linked List as the data structure for our analysis. In MTF algorithm the reorganization cost is considered to be zero as there is no paid exchange.

**Theorem 1**- *Let $C_{MTF}(\ell, T_1)$ be the total access cost incurred by MTF algorithm while serving a request sequence $T_1$ on a list $\ell$ of size $n$ then $C_{MTF}(\ell, T_1) = [n^2 \times (2k-1)+n]/2$.*

**Proof:** Let $C_{MTF}(\ell, T_1)$ be the total access cost incurred by MTF algorithm while serving a request sequence $T_1$ on a given list $\ell$ with initial list configuration $<\ell_1, \ell_2, \ldots, \ell_n>$. Let $T_{1i}$ be a subsequence of $T_1$ for $i = 1, 2, 3, \ldots, k$. So $T_1 = T_{11}T_{12}T_{13}\ldots T_{1k}$ where each $T_{1i} = \Pi_\ell = <\ell_1, \ell_2, \ell_3, \ldots, \ell_n>$ for $i=1, 2, 3, \ldots, k$. Let $C_{MTF}(L_i, T_{1i})$ be the total access cost of serving a request subsequence $T_{1i}$ of $T_1$ on a list configuration $L_i$. Here $L_i$ denotes a configuration of the list $\ell$ before serving the subsequence $T_{1i}$. The total access cost $C_{MTF}(\ell, T_1)$ can be calculated as $C_{MTF}(\ell, T_1) = C_{MTF}(L_1, T_{11}) + C_{MTF}(L_2, T_{12}) + \ldots C_{MTF}(L_k, T_{1k}) = \sum_{i=1}^{k} C_{MTF}(L_i, T_{1i})$

*Step 1:* **Computation of $C_{MTF}(L_i, T_{1i})$ for $i=1$**

Let $\sigma$ be a type $T_{11}$ request subsequence of $T_1$ that is served with list configuration $L_1 = \ell = <\ell_1, \ell_2, \ell_3, \ldots, \ell_n>$. Let $\sigma_j$ be the $j^{th}$ request of the request subsequence $\sigma$ and $C_{\sigma j}(MTF)$ denotes the access cost of serving a request $\sigma_j$ for $j=1, 2, 3, \ldots, n$ using MTF algorithm. So as shown in figure 1, $C_{\sigma 1}(MTF) = 1, C_{\sigma 2}(MTF) = 2, C_{\sigma 3}(MTF) = 3 \ldots C_{\sigma n}(MTF) = n$. Hence $C_{MTF}(L_1, T_{11}) = \sum_{j=1}^{n} C_{\sigma j}(MTF) = 1+2+3+\ldots+n = n(n+1)/2$.

*Step2:* **Computation of $C_{MTF}(L_i, T_{1i})$ for $i=2, 3, \ldots, k$**

As shown in figure 1, $C_{MTF}(L_i, T_{1i}) = n + n + \ldots$(n times) for each $i = 2, 3, \ldots, k$. Hence $\sum_{i=2}^{k} = C_{MTF}(L_i, T_{1i}) = (k-1) \times n^2$

*Step3:* **Computation of $C_{MTF}(\ell, T_1)$**

$C_{MTF}(\ell, T_1) = C_{MTF}(L_1, T_{11}) + \sum_{i=2}^{k} C_{MTF}(L_i, T_{1i})$
$= [n(n+1)]/2 + (k-1) \times n^2 = [n^2 \times (2k-1)+n]/2$

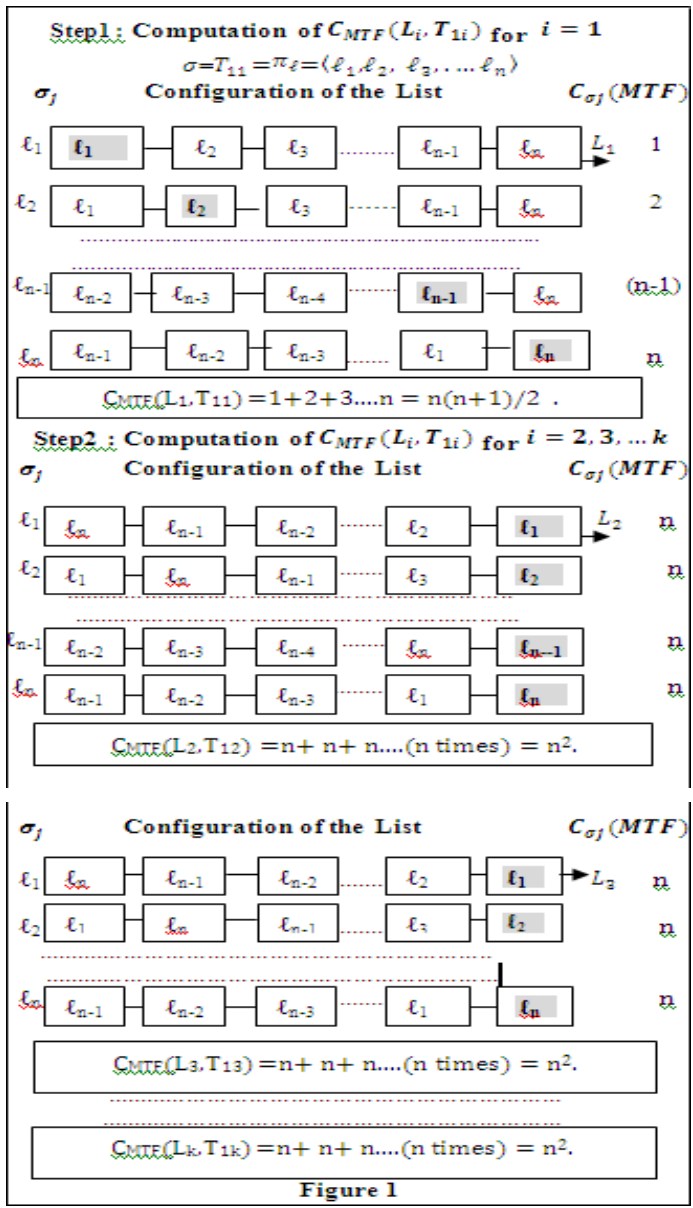

Figure 1

**Theorem 2-** Let $C_{MTF}(\ell, T_2)$ be the total access cost incurred by MTF algorithm while serving a request sequence $T_2$ on a list $\ell$ of size n then $C_{MTF}(\ell, T_2) = k \times n^2$.

**Proof:** Let $C_{MTF}(\ell, T_2)$ be the total access cost incurred by MTF algorithm while serving a request sequence $T_2$ on a given list $\ell$ with initial list configuration $<\ell_1, \ell_2, \ell_3, \ldots, \ell_n>$. Let $T_{2i}$ be a subsequence of $T_2$ for $i=1, 2, 3\ldots, k$. So $T_2 = T_{21}T_{22}T_{23}\ldots T_{2k}$ where each $T_{2i} = \Pi_r = <\ell_n, \ell_{n-1}\ldots, \ell_2, \ell_1>$ for $i=1, 2, 3\ldots, k$. Let $C_{MTF}(L_i, T_{2i})$ be the total access cost of serving a request subsequence $T_{2i}$ of $T_2$ on a list configuration $L_i$. Here $L_i$ denotes a configuration of the list $\ell$ before serving the subsequence $T_{2i}$ for $i =1, 2, 3\ldots, k$. The total access cost $C_{MTF}(\ell, T_2)$ can be calculated a $C_{MTF}(\ell, T_2) = C_{MTF}(L_1, T_{21}) + C_{MTF}(L_2, T_{22}) +\ldots C_{MTF}(L_k, T_{2k}) = \sum_{i=1}^{k} C_{MTF}(L_i, T_{2i})$

**Step 1:** Computation of $C_{MTF}(L_i, T_{2i})$ for i =1
Let σ be a type $T_{21}$ request subsequence of $T_2$ that is served with list configuration $L_1 = \ell = <\ell_1, \ell_2, \ell_3\ldots \ell_n>$. Let $\sigma_j$ be the jth request of the request subsequence σ and $C_{\sigma j}(MTF)$ denotes the access cost of serving a request $\sigma_j$ for $j=1, 2,..n$ using MTF algorithm. $C_{\sigma 1}(MTF) = n$, $C_{\sigma 2}(MTF) = n, \ldots C_{\sigma n}(MTF) = n$.

Hence $C_{MTF}(L_1, T_{21}) = \sum_{j=1}^{n} C_{\sigma j}(MTF) = n + n +\ldots(n\ times) = n^2$.

**Step2:** Computation of $C_{MTF}(L_i, T_{2i})$ for i = 2, 3, 4\ldots k
$C_{MTF}(L_i, T_{2i}) = n + n +\ldots (n\ times) = n^2$ for each i = 2, 3,\ldots, k. Hence $\sum_{i=2}^{k} C_{MTF}(L_i, T_{2i}) = (k-1)n^2$

**Step3:** Computation of $C_{MTF}(\ell, T_2)$
$C_{MTF}(\ell, T_2) = C_{MTF}(L_1, T_{21}) + \sum_{i=2}^{k} C_{MTF}(L_i, T_{2i}) = n^2 + (k-1)n^2 = k \times n^2$

**Theorem 3–** Let $C_{MTF}(\ell, T_3)$ be the total access cost incurred by MTF algorithm while serving a request sequence $T_3$ on a list $\ell$ of size n then $C_{MTF}(\ell, T_3) = \sum_{j=1}^{n} P_j + (k-1)n^2$ where $P_j$ denotes the position of $\sigma_j$ before accessing $\sigma_j$ in the list for j=1, 2, 3\ldots n.

**Proof:** Let $C_{MTF}(\ell, T_3)$ be the total access cost incurred by MTF algorithm while serving a request sequence of type $T_3$ on a given list $\ell$ with initial list configuration $<\ell_1, \ell_2, \ell_3\ldots \ell_n>$. Let $T_{3i}$ be a subsequence of $T_3$ for i =1, 2, 3\ldots, k. So $T_3 = T_{31}T_{32}T_{33}\ldots T_{3k}$ where each $T_{3i} = \Pi_x$ be a particular permutation of $\ell$ except $\Pi_\ell$ and $\Pi_r$ for i=1, 2, 3\ldots, k. Let $C_{MTF}(L_i, T_{3i})$ be the total access cost of serving a request subsequence $T_{3i}$ of $T_3$ on a list configuration $L_i$. Here $L_i$ denotes a configuration of the list $\ell$ before serving the subsequence $T_{3i}$ for i=1, 2, 3\ldots k. The total access cost $C_{MTF}(\ell, T_3)$ can be calculated as $C_{MTF}(\ell, T_3) = C_{MTF}(L_1, T_{31}) + C_{MTF}(L_2, T_{32}) +\ldots C_{MTF}(L_k, T_{3k}) = \sum_{i=1}^{k} C_{MTF}(L_i, T_{3i})$

**Step 1:** Computation of $C_{MTF}(L_i, T_{3i})$ for i =1
Let σ be a type $T_{31}$ request subsequence of $T_3$ that is served with list configuration $L_1 = \ell = <\ell_1, \ell_2, \ell_3\ldots \ell_n>$. Let $\sigma_j$ be the jth request of the request subsequence σ and $C_{\sigma j}(MTF)$ denotes the access cost of serving a request $\sigma_j$ for j=1, 2, 3,\ldots, n using MTF algorithm. Let $P_j$ denotes the position of $\sigma_j$ before accessing $\sigma_j$ in the list for j=1, 2, 3\ldots n. Hence $C_{\sigma 1}(MTF) = P_1$, $C_{\sigma 2}(MTF) = P_2$, $C_{\sigma 3}(MTF) = P_3,\ldots C_{\sigma n}(MTF) = P_n$. Hence $C_{MTF}(L_1, T_{31}) = \sum_{j=1}^{n} C_{\sigma j}(MTF) = P_1 + P_2 + P_3\ldots+ P_n = \sum_{j=1}^{n} P_j$

**Step2:** Computation of $C_{MTF}(L_i, T_{3i})$ for i =2, 3,\ldots, k
$C_{MTF}(L_i, T_{3i}) = n + n +\ldots (n\ times) = n^2$ for each i = 2, 3,\ldots, k. Hence $\sum_{i=2}^{k} C_{MTF}(L_i, T_{3i}) = (k-1) \times n^2$

**Step3:** Computation of $C_{MTF}(\ell, T_3)$
$C_{MTF}(\ell, T_3) = C_{MTF}(L_1, T_{31}) + \sum_{i=2}^{k} C_{MTF}(L_i, T_{3i}) = \sum_{j=1}^{n} P_j + (k-1)n^2$

**Theorem 4-** Let $C_{MTF}(\ell, T_4)$ be the total access cost incurred by MTF algorithm while serving a request sequence $T_4$ on a list $\ell$ of size n then $C_{MTF}(\ell, T_4) = \sum_{j=1}^{q} P_j + (k-1)q^2$ where $P_j$ denotes the position of $\sigma_j$ before accessing $\sigma_j$ in the list for j=1, 2, 3\ldots q.

**Proof:** Let $C_{MTF}(\ell, T_4)$ be the total access cost incurred by MTF algorithm while serving a request sequence of type $T_4$ on a given list $\ell$ with initial list configuration $<\ell_1, \ell_2, \ell_3\ldots \ell_n>$. Let $T_{4i}$ be a subsequence of $T_4$ for i = 1, 2, 3\ldots k. So $T_4 = T_{41}T_{42}T_{43}\ldots T_{4k}$ where each $T_{4i} = \delta_x$ be a particular subsequence of the list $\ell$ of size q for i =1, 2, 3\ldots k. Let $C_{MTF}(L_i, T_{4i})$ be the total access cost of serving a request subsequence $T_{4i}$ of $T_4$ on a list configuration $L_i$. Here $L_i$ denotes a configuration of the list $\ell$ before serving the subsequence $T_{4i}$ for i =1, 2, 3\ldots, k. The total access cost $C_{MTF}(\ell, T_4)$ can be calculated as $C_{MTF}(\ell, T_4) = C_{MTF}(L_1, T_{41}) + C_{MTF}(L_2, T_{42}) + \ldots C_{MTF}(L_k, T_{4k}) = \sum_{i=1}^{k} C_{MTF}(L_i, T_{4i})$

***Step 1:*** **Computation of $C_{MTF}(L_i, T_{4i})$ for i =1**

Let σ be a type $T_{41}$ request subsequence of $T_4$ that is served with list configuration $L_1 = \ell = <\ell_1, \ell_2, \ell_3..., \ell_n>$. Let $\sigma_j$ be the $j^{th}$ request of the request subsequence σ and $C_{\sigma j}(MTF)$ denotes the access cost of serving a request $\sigma_j$ for j=1, 2, 3…., q using MTF algorithm. Let $P_j$ denotes the position of $\sigma_j$ before accessing $\sigma_j$ in the list for j=1, 2, 3….q. Hence $C_{\sigma 1}(MTF) = P_1$, $C_{\sigma 2}(MTF) = P_2$, $C_{\sigma 3}(MTF) = P_3$.... $C_{\sigma q}(MTF = P_q$. Hence $C_{MTF}(L_1, T_{41}) = \sum_{j=1}^{n} C_{\sigma j}(MTF) = P_1 + P_2 + P_3 ….+ P_n = \sum_{j=1}^{q} P_j$

***Step2:*** **Computation of $C_{MTF}(L_i, T_{4i})$ for i =2, 3...., k**

$C_{MTF}(L_i, T_{4i}) = q + q +$..... (q times) $= q^2$ for each i = 2, 3…., k. Hence $\sum_{i=2}^{k} C_{MTF}(L_i, T_{4i}) = (k-1) \times q^2$

***Step3:*** **Computation of $C_{MTF}(\ell, T_4)$**

$C_{MTF}(\ell, T_4) = C_{MTF}(L_1, T_{41}) + \sum_{i=2}^{k} C_{MTF}(L_i, T_{4i}) = \sum_{j=1}^{q} P_j + (k-1)q^2$

Table 1. Summary of our results

| Request sequence Type | Total Access Cost |
|---|---|
| $T_1$ | $[n^2 \times (2k-1)+n]/2$ |
| $T_2$ | $k \times n^2$ |
| $T_3$ | $\sum_{j=1}^{n} P_j + (k-1)n^2$ |
| $T_4$ | $\sum_{j=1}^{n} P_j + (k-1)q^2$ |

*c) Graphical Representation of Results*

We have compared the performance of MTF algorithm for only two types of request sequences i.e. $T_1$ and $T_2$. As evident from our theoretical results, the access cost of MTF for $T_3$ will vary between $T_1$ and $T_2$ and sometimes this value will closely match with either $T_1$ or $T_2$. The access cost for $T_4$ will depend on the size of the subsequence q, and the number of times the subsequence is repeated. Hence the performance of MTF for $T_1$ and $T_2$ can be compared. Our comparison results are graphically shown in figure 2, figure3 and figure 4. Let C1= $C_{MTF}(\ell, T_1)$ and C2= $C_{MTF}(\ell, T_2)$. In figure 2, keeping the value of k constant, we plot a graph by taking values of n in x-axis and total access costs C1 and C2 in y-axis. In figure 3, keeping the value of n constant we plot a graph by taking values of k in x-axis and total access costs C1 and C2 in y-axis. In figure 4, we plot a graph by taking the values of n and k in increasing order in x-axis and the total access costs C1 and C2 in y-axis.

### III. CONCLUSION AND FUTURE WORK

In this paper we have considered four special types of request sequences without locality of reference corresponding to real life inputs. Using these request sequences, we have analyzed the performance of MTF algorithm and have obtained some novel and interesting theoretical results for computing the total access cost. We have observed that MTF algorithm performs better for $T_1$ type of request sequence than for $T_2$ type of request sequence. However for request sequence $T_3$ the performance of MTF will vary between that of $T_1$ and $T_2$. The performance of MTF for $T_4$ will depend on the size of the subsequence and the no of times the subsequence is repeated.

More special types of request sequences can be generated for real life inputs. Performance evaluation of MTF and other of List accessing algorithms can be done for the same types of request sequences as a future work.

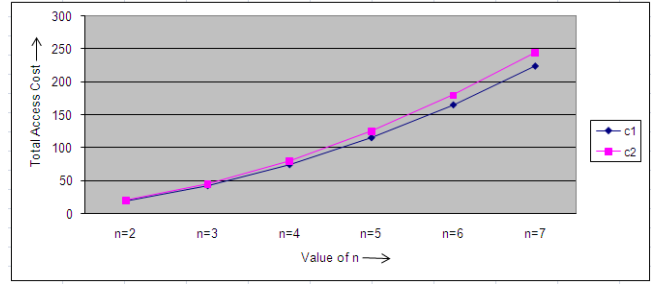

Figure 2    For constant k (k = 5)

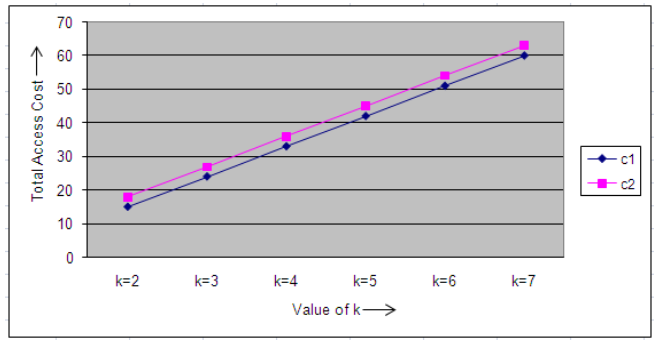

Figure 3  For constant n (n = 3)

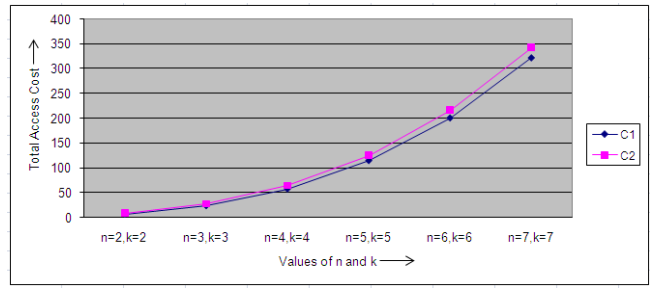

Figure 4    For increasing n and k